\documentclass[sn-vancouver]{sn-jnl}

\jyear{2021}%

\theoremstyle{thmstyleone}%
\theoremstyle{thmstyletwo}%

\theoremstyle{thmstylethree}%
\newcommand{\package}{\emph{RamanSPy}}

\usepackage{listings}
\usepackage{xcolor}

\definecolor{codegreen}{rgb}{0,0.6,0}
\definecolor{codegray}{rgb}{0.5,0.5,0.5}
\definecolor{codepurple}{rgb}{0.58,0,0.82}
\definecolor{backcolour}{rgb}{0.95,0.95,0.92}

\lstdefinestyle{mystyle}{
    backgroundcolor=\color{backcolour},   
    commentstyle=\color{codegreen},
    keywordstyle=\color{magenta},
    numberstyle=\tiny\color{codegray},
    stringstyle=\color{codepurple},
    basicstyle=\ttfamily\footnotesize,
    breakatwhitespace=false,         
    breaklines=true,                 
    captionpos=b,                    
    keepspaces=true,                 
    numbers=left,                    
    numbersep=5pt,                  
    showspaces=false,                
    showstringspaces=false,
    showtabs=false,                  
    tabsize=2
}

\begin{document}


\title{\package{}: An open-source Python package for integrative Raman spectroscopy data analysis}


\author[1]{\fnm{Dimitar} \sur{Georgiev}}

\author[2,3]{\fnm{Simon} \sur{Vilms Pedersen}}

\author[2]{\fnm{Ruoxiao} \sur{Xie}}

\author[2]{\fnm{Álvaro} \sur{Fernández-Galiana}}

\author*[2]{\fnm{Molly M.} \sur{Stevens}\email{m.stevens@imperial.ac.uk}}

\author*[4]{\fnm{Mauricio} \sur{Barahona}\email{m.barahona@imperial.ac.uk}}

\affil[1]{\orgdiv{Department of Computing}, \orgname{Imperial College London}, \orgaddress{
\country{United Kingdom}}}

\affil[2]{\orgdiv{Department of Materials,  Department of Bioengineering \& Institute of Biomedical Engineering}, \orgname{Imperial College London}, \orgaddress{
\country{United Kingdom}}}

\affil[3]{Present address: \orgname{University of Southern Denmark}, \orgaddress{
\city{Odense}, 
\country{Denmark}}}

\affil[4]{\orgdiv{Department of Mathematics}, \orgname{Imperial College London}, \orgaddress{
\country{United Kingdom}}}

\abstract{Raman spectroscopy is a non-destructive and label-free chemical analysis technique, which plays a key role in the analysis and discovery cycle of various branches of science. Nonetheless, progress in Raman spectroscopic analysis is still impeded by the lack of software, methodological and data standardisation, and the ensuing fragmentation and lack of reproducibility of analysis workflows thereof. To address these issues, we introduce \package{}, an open-source Python package for Raman spectroscopic research and analysis. \package{} provides a comprehensive library of ready-to-use tools for spectroscopic analysis, which streamlines day-to-day tasks, integrative analyses, as well as novel research and algorithmic development. \package{} is modular and open source, not tied to a particular technology or data format, and can be readily interfaced with the burgeoning ecosystem for data science, statistical analysis and machine learning in Python. }


\maketitle


Raman spectroscopy (RS) is a powerful sensing modality based on inelastic light scattering, which provides qualitative and quantitative chemical analysis with high sensitivity and specificity \cite{colthup2012introduction}. RS yields a characterisation of the vibrational profile of molecules, which can help elucidate the composition of chemical compounds, biological specimens and materials \cite{mccreery2005raman, shipp2017raman, Fernandez-Galiana:2023:10.1002/adma.202210807}. In contrast to most conventional technologies for (bio)chemical characterisation (e.g., staining, different omics, fluorescence microscopy and mass spectrometry), RS is both label-free and non-destructive, thereby allowing the acquisition of rich biological and chemical information without compromising the structural and functional integrity of the probed samples. This advantage has enabled a broad range of applications of RS in biomedical and pharmaceutical research \cite{butler2016using, movasaghi2007raman, vankeirsbilck2002applications}, including in the imaging of cells and tissues \cite{smith2016raman, kallepitis2017quantitative, Lalone:2023, pedersen2023spectral}, the chemical analysis of drug compounds \cite{wang2018research, VANKEIRSBILCK2002869}, and the detection of disease \cite{auner2018applications, mahadevan1996raman, kong2015raman, penders2021single}. 

An area of topical interest is the frontier of Raman spectroscopy, chemometrics and artificial intelligence (AI), with its promise of more autonomous, flexible and data-driven RS analytics \cite{pan2022review, luo2022deep, lussier2020deep}. There has been a recent surge in the adoption of AI methods in Raman-based research~\cite{Fernandez-Galiana:2023:10.1002/adma.202210807}, with applications to RS now spanning domains as broad as the identification of pathogens and other microbes \cite{bacteria, lu2020combination, yan2021raman, wang2021applications}; the characterisation of chemicals, including minerals \cite{carey2015machine}, pesticides \cite{zhu2021rapid} and other analytes \cite{han2020bayesian, akpolat2020high}; the development of novel diagnostic platforms \cite{ralbovsky2020towards, talari2019advancing, heng2021advances, zhang2022raman}; as well as the application of techniques from computer vision for denoising and super-resolution in Raman imaging \cite{deeper}.

As new hardware, software and data acquisition RS technologies continue to emerge~\cite{qi2023recent, zhu2014technical}, there is a pressing need for an integrated RS data analysis environment, which facilitates the development of pipelines, methods and applications, and bolsters the use of RS in biomedical research. Yet, the full deployment of RS and its capabilities is still hindered by practical factors stemming from the restrictive, functionally disparate, and highly encapsulated nature of current commercial software for RS data analysis. 
RS data analysis often operates within proprietary software environments and data formats, which have induced methodological inconsistencies and reduced cross-platform and benchmarking efforts, with growing concerns around reproducibility. These restrictions have also 
hampered the adoption of new AI technologies into the field~\cite{byrne2016spectral, tanwar2021advancing, barton2022chemometrics, moller2017robust, ntziouni2022review}. As a consequence, researchers increasingly resort to developing in-house scripts for RS analysis in Python \cite{van1995python}, further adding to methodological fragmentation and lack of standardisation~\cite{guo2021chemometric}. 

In response to these challenges, we have developed \package{} - a modular, open-source framework for integrated \underline{Raman} \underline{S}pectroscopy analytics in \underline{Py}thon. \package{} is designed to systematise day-to-day workflows, enhance algorithmic development and validation, and accelerate the adoption of novel AI technologies into the RS field. 
Firstly, \package{} serves as a platform for general-purpose RS analytics supporting the RS data life cycle by providing a suite of ready-to-use modules for data loading, preprocessing, analysis and visualisation. By design, these functionalities are not tied to any specific technology or data type, thereby allowing integrative and transferable cross-platform analyses.
Secondly, \package{} addresses challenges in data preprocessing by facilitating the compilation of reproducible pipelines to streamline and automatise preprocessing protocols. 
Thirdly, \package{} helps bridge the gap between RS data and state-of-the-art AI technologies within the extensive machine learning (ML) ecosystem in Python. Complemented by direct access to Raman datasets, preprocessing protocols and performance metrics, this provides the foundation for AI model development and benchmarking.

The codebase of \package{} is hosted at \url{https://github.com/barahona-research-group/RamanSPy} with extended documentation (\url{https://ramanspy.readthedocs.io}), which includes tutorials and example applications, and details about the real-world research applications presented in this paper.

\begin{figure}[t]
    \centering
    \includegraphics[width=\textwidth]{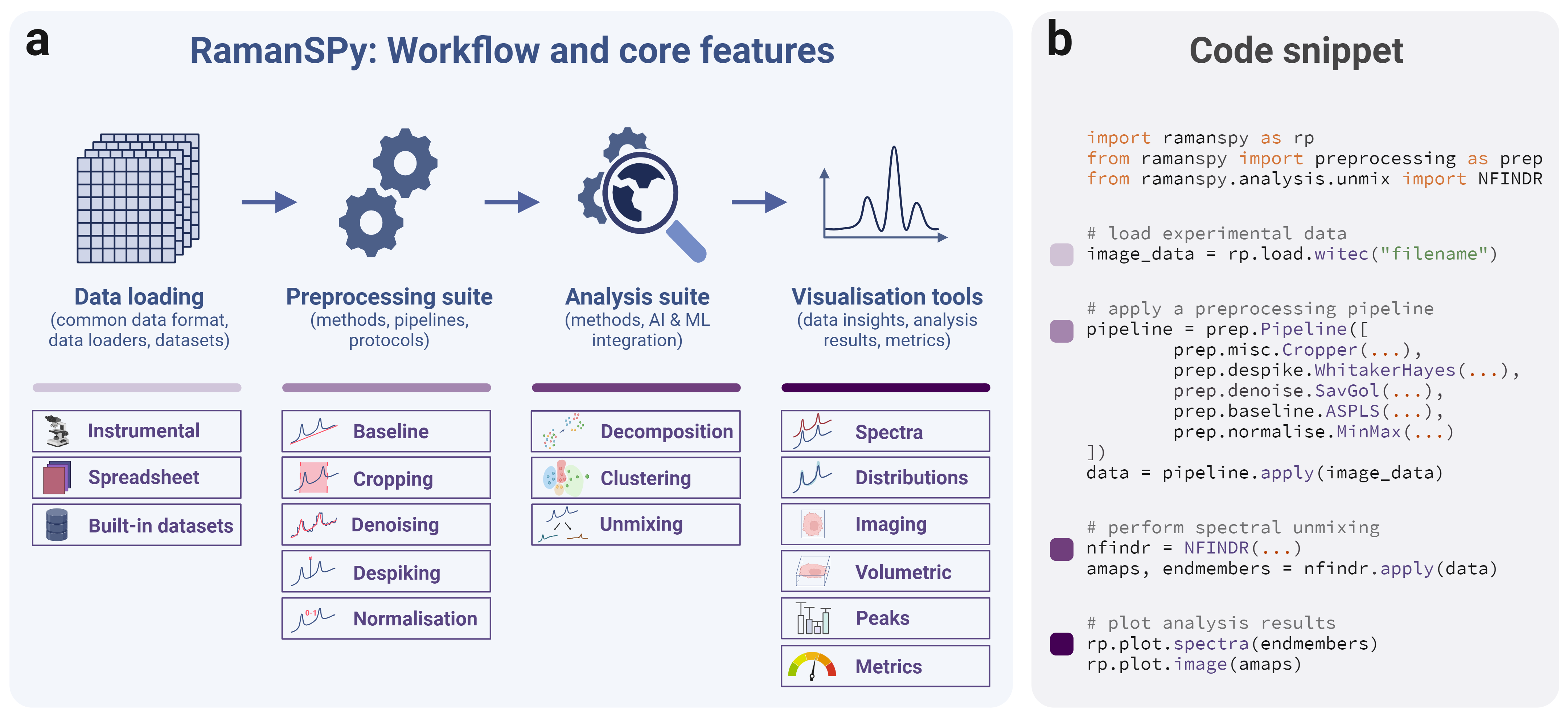}
    \caption{\textbf{General Raman spectroscopy workflow and core features of \package{}.} \textbf{a,} \package{} supports the Raman spectroscopic data analysis life cycle via a modular, loosely coupled architecture. RS data is parsed to a common data representation format, which is interfaced with preprocessing, analysis and visualisation tools within \package{}. 
    The core features of \package{} include a comprehensive library of standardised, simple-to-use procedures for data loading, preprocessing, analysis and visualisation. These modules are flexible and allow the incorporation of further techniques and in-house methods. For complete information about the modules available in \package{}, refer to the documentation at \url{https://ramanspy.readthedocs.io}.
    \textbf{b,} An example workflow use case in \package{}: Raman data is loaded, preprocessed and analysed in a few lines of code. 
    }
    \label{fig:overview}
\end{figure}

\section*{Results}
\bmhead{\package{} as a platform for general Raman spectroscopy analytics}

\package{} is based on a modular, object-oriented programming (OOP) infrastructure, which streamlines the RS data analysis life cycle (Fig.~\ref{fig:overview}a) and allows users to compile diverse analysis workflows with a few lines of reusable, user-friendly code (Fig. \ref{fig:overview}b). The framework adopts a scalable array-based data representation, which accommodates different spectroscopic modalities, including single-point spectra, Raman imaging data, and volumetric scans. Experimental data can be loaded through custom loaders built into \package{} or through standard tools available in Python. The data representation functions as a common data container that defines the interface between RS data management and manipulation within \package{}, allowing us to unify data standards across setups and vendors, independent of instrumental origin and acquisition modality.

\package{} also provides an extensive toolbox for preprocessing, analysis and visualisation. The preprocessing suite includes techniques for denoising, baseline correction, cosmic spike removal, normalisation and background subtraction, among others. Likewise, the analysis toolbox includes modules for decomposition (useful for dimensionality reduction), clustering and spectral unmixing. \package{} also includes a set of data visualisation tools.
All these modules are organised into an extensible class structure, which standardises their application across projects and datasets to facilitate transferable analysis workflows.  

\begin{figure}
    \centering
    \includegraphics[width=\textwidth]{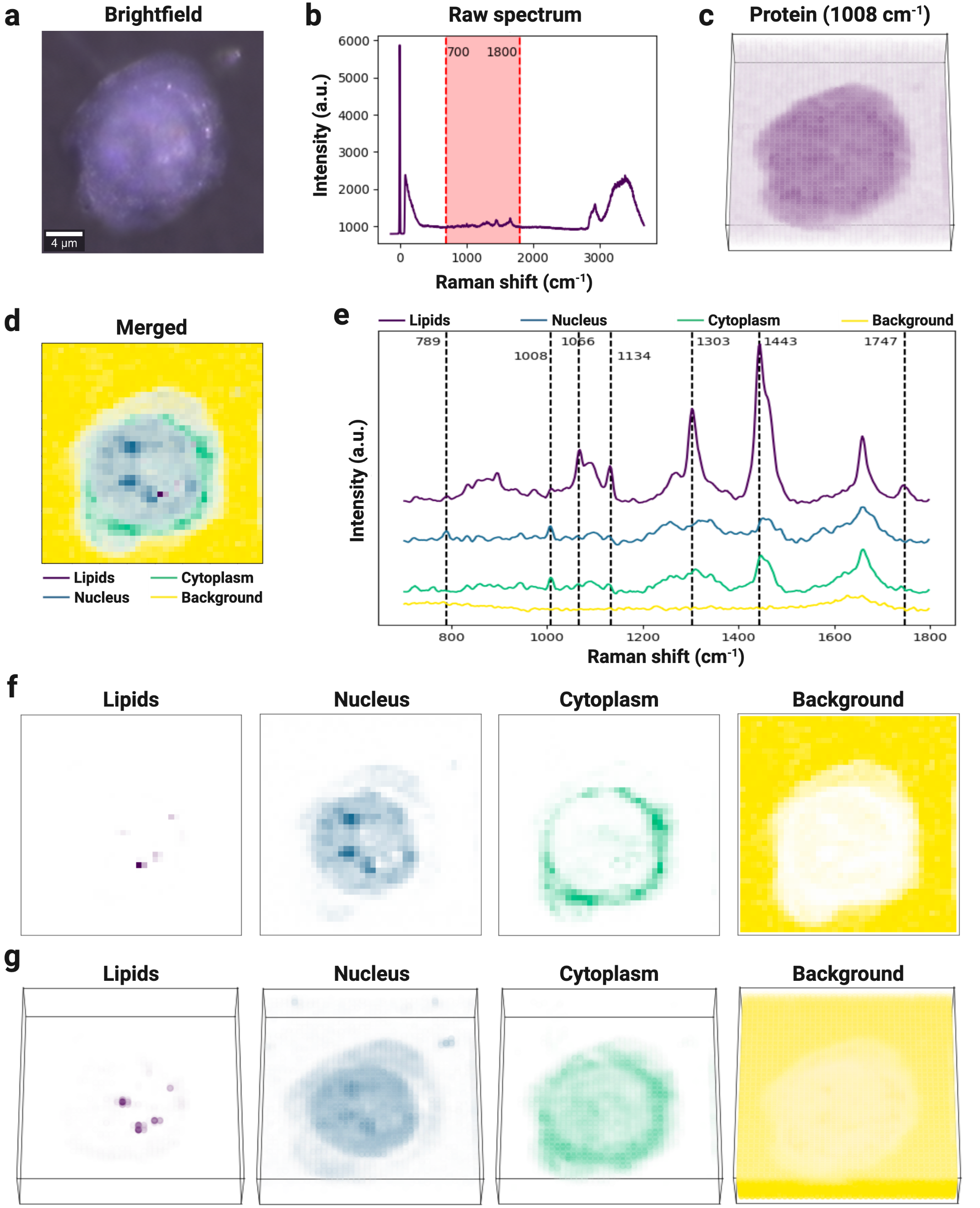} 
    \caption{\textbf{Morphological analysis of a THP-1 cell via spectral unmixing with \package{}.} \textbf{a,} Bright-field image of a THP-1 cell. The same cell was also imaged with Raman spectroscopy. Image and volumetric Raman data from \cite{kallepitis2017quantitative}. \textbf{b,} An exemplar spectrum from the raw volumetric Raman data (taken from the centre of the layer in \textbf{d}). The fingerprint region (700--1800 cm$^{-1}$) shaded in red was used for the analysis. \textbf{c,} Volumetric data at the 1008 cm$^{-1}$ band (characteristic of proteins) after preprocessing. \textbf{d-g,} Spectral unmixing analysis reveals the distribution of components within the cell: lipids (violet), nucleus (blue), cytoplasm (green), and background (yellow). 
    \textbf{d,} A merged reconstruction of the sixth depth layer (10 in total) of the THP-1 cell determined via spectral unmixing. \textbf{e,} Four endmembers derived with N-FINDR \cite{nfindr} characterised via peak assignment. \textbf{f,} Fractional abundance maps calculated with FCLS \cite{fcls} for the sixth depth layer. \textbf{g,} Fractional abundance maps for the entire volume. 
    }
    \label{fig:core_example}
\end{figure}

We showcase the core features of \package{} by 
analysing volumetric Raman spectroscopic data from a human leukaemia monocytic (THP-1) cell~\cite{kallepitis2017quantitative} (Fig. \ref{fig:core_example}). The aim is to investigate the cell phenotype in a label-free manner using RS and methods from chemometrics. We load the data using built-in tools, and perform a spectral preprocessing protocol comprising spectral cropping to the fingerprint region (700--1800 cm$^{-1}$), cosmic spike removal, denoising, baseline correction and normalisation (see SI). 
Using the visualisation tools in the package, we inspect data quality (Fig. \ref{fig:core_example}b) and perform initial exploratory analysis by examining, e.g., data slices across wavenumber bands (Fig. \ref{fig:core_example}c). The analysis proceeds to spectral unmixing based on: (i)  N-FINDR \cite{nfindr} for endmember detection, and (ii) fully constrained least squares (FCLS) \cite{fcls} for component quantification. This process is exploited to demix signal contributions from different cellular components and study their morphological organisation within the THP-1 cell. Following the peak assignment in \cite{kallepitis2017quantitative}, we distinguish endmember components related to lipids (band 1008 cm$^{-1}$), nucleic acid (band 789 cm$^{-1}$), cytoplasm (bands 1066, 1134, 1303, 1443 and 1747 cm$^{-1}$), and the background (Fig. \ref{fig:core_example}e). Finally, we produce fractional abundance reconstructions based on the extracted endmembers, which we can examine on a single-layer level (Fig. \ref{fig:core_example}f) and across the entire volume (Fig. \ref{fig:core_example}g) to localise cellular organelles within the cell.

\bmhead{\package{} enables automated pipelining of spectral preprocessing protocols} 

Experimental RS data is susceptible to non-specific signal artefacts (e.g., cosmic rays, autofluorescence background, variability in instrumentation), which can severely affect downstream analyses. Preprocessing is therefore a critical step in any spectroscopic analysis workflow \cite{gautam2015review, ryabchykov2018analyzing}. 
Yet, due to a lack of standardisation and frameworks for general-purpose pipelining~\cite{byrne2016spectral}, researchers tend to utilise variable preprocessing protocols, often dispersed across different software systems, thus affecting reproducibility and validation \cite{rozenstein2014comparing, alshdaifat2021effect}.

\begin{figure}
    \centering
    \includegraphics[width=1\textwidth]{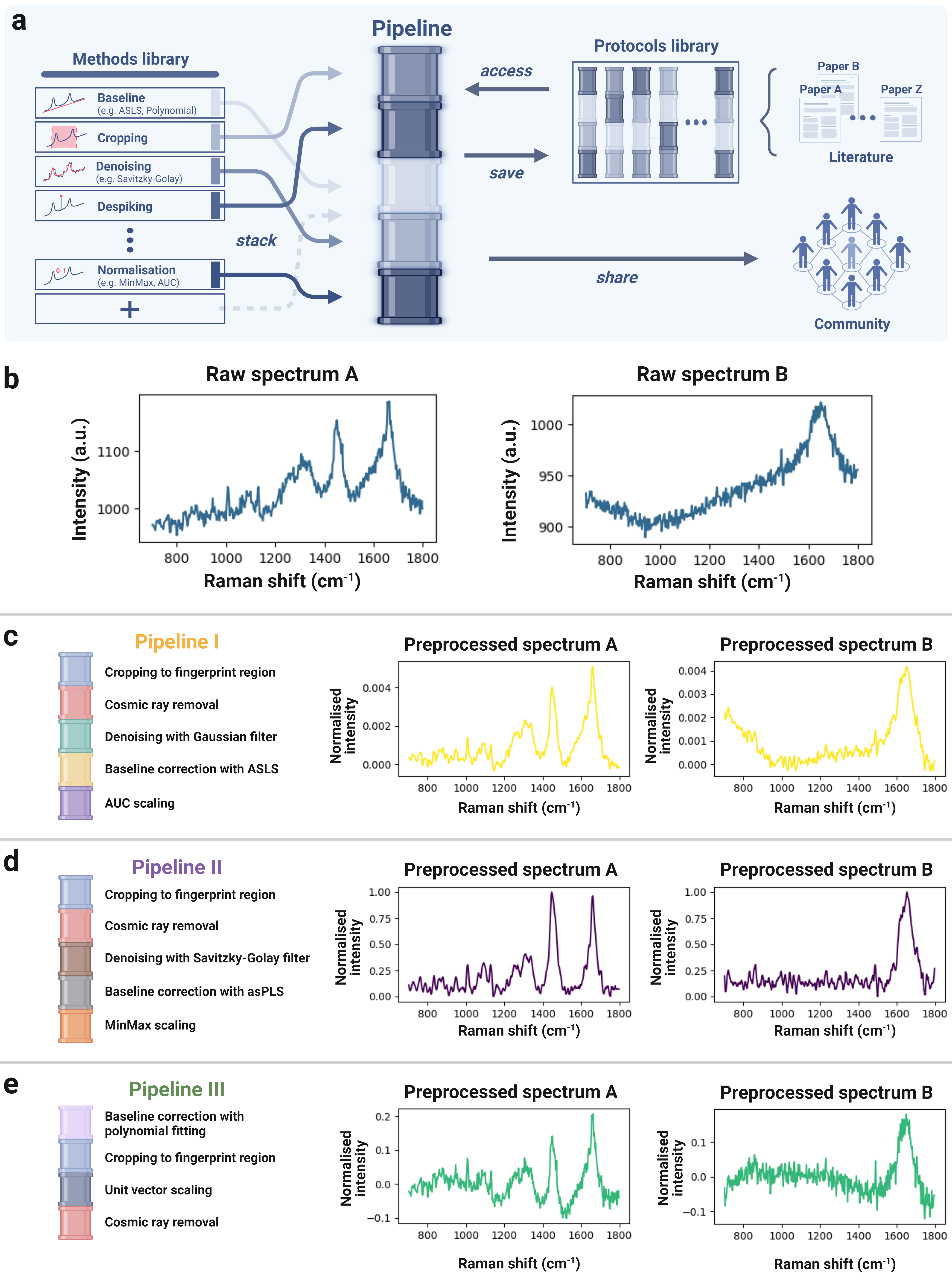}
    \caption{\textbf{Spectral preprocessing pipelining in \package{}.} 
    \textbf{a, } \package{} automates the construction, customisation and execution of multi-layered preprocessing procedures via pipelining. Users can assemble built-in and in-house methods into complete preprocessing pipelines, which are fully compatible with data integrated within \package{} and can be saved, reused and shared. \package{} also provides access to a library of already assembled preprocessing pipelines. \textbf{b, } Two raw spectra from the THP-1 data from \cite{kallepitis2017quantitative} are used to compare the effect of different preprocessing pipelines. \textbf{c-e, } The results of three preprocessing pipelines built within \package{}, demonstrating the need for standardisation. Note on preprocessing methods: fingerprint region is $700$--$1800$ cm$^{-1}$; ASLS - Asymmetric Least Squares \cite{eilers2005baseline};  asPLS - Adaptive Smoothness Penalized Least Squares \cite{zhang2020baseline}; AUC - area under the curve; cosmic rays removed with algorithm from \cite{whitaker2018simple}. 
    }
    \label{fig:preprocessing_pipelines}
\end{figure}

To facilitate the creation of reproducible protocols, \package{} incorporates a pipelining infrastructure, which systematises the process of creating, customising and executing preprocessing pipelines (Fig.~\ref{fig:preprocessing_pipelines}a). Users can use a specialised class, which defines a generic, multi-layered preprocessing procedure, to assemble pipelines from selected built-in preprocessing modules or other in-house methods. 
To reduce overhead, the constructed pipelines are designed to function exactly as any single method, i.e., they are fully compatible with the rest of the modules and data structures in the package. Furthermore, pipelines can be easily saved, reused and shared to foster the development of a repository of preprocessing protocols. As a seed to this repository, \package{} provides a library of assembled preprocessing protocols (custom pre-defined, or adapted from the literature \cite{bergholt2016raman}), which users can access and exploit. 

To illustrate the pipelining functionalities, we use \package{} to construct three preprocessing protocols
by compiling selected methods in the desired order of execution, and applying them out-of-the-box to data loaded into the platform (Fig. \ref{fig:preprocessing_pipelines}c-e). We use them to preprocess Raman spectroscopic data from \cite{kallepitis2017quantitative} (Fig. \ref{fig:preprocessing_pipelines}b). 
Note how the three pipelines yield substantially different results, reinforcing the importance of consistency in the selection of preprocessing protocols. Pipeline II was deemed the most robust, and consequently added to the protocols library in \package{} as default.

\bmhead{\package{} facilitates AI integration and validation of next-generation Raman data analytics}

To help accelerate the adoption of AI technologies for RS analysis, \package{} is endowed with a permeable architecture, which streamlines the interface between Raman spectroscopic data and the burgeoning ML ecosystem in Python. This is complemented by tools for benchmarking, such as datasets and performance metrics, which support the evaluation of new models and algorithms. We show below two examples of \package{}'s capabilities for ML integration and benchmarking.

\begin{figure}
    \centering
    \includegraphics[width=1 \textwidth]{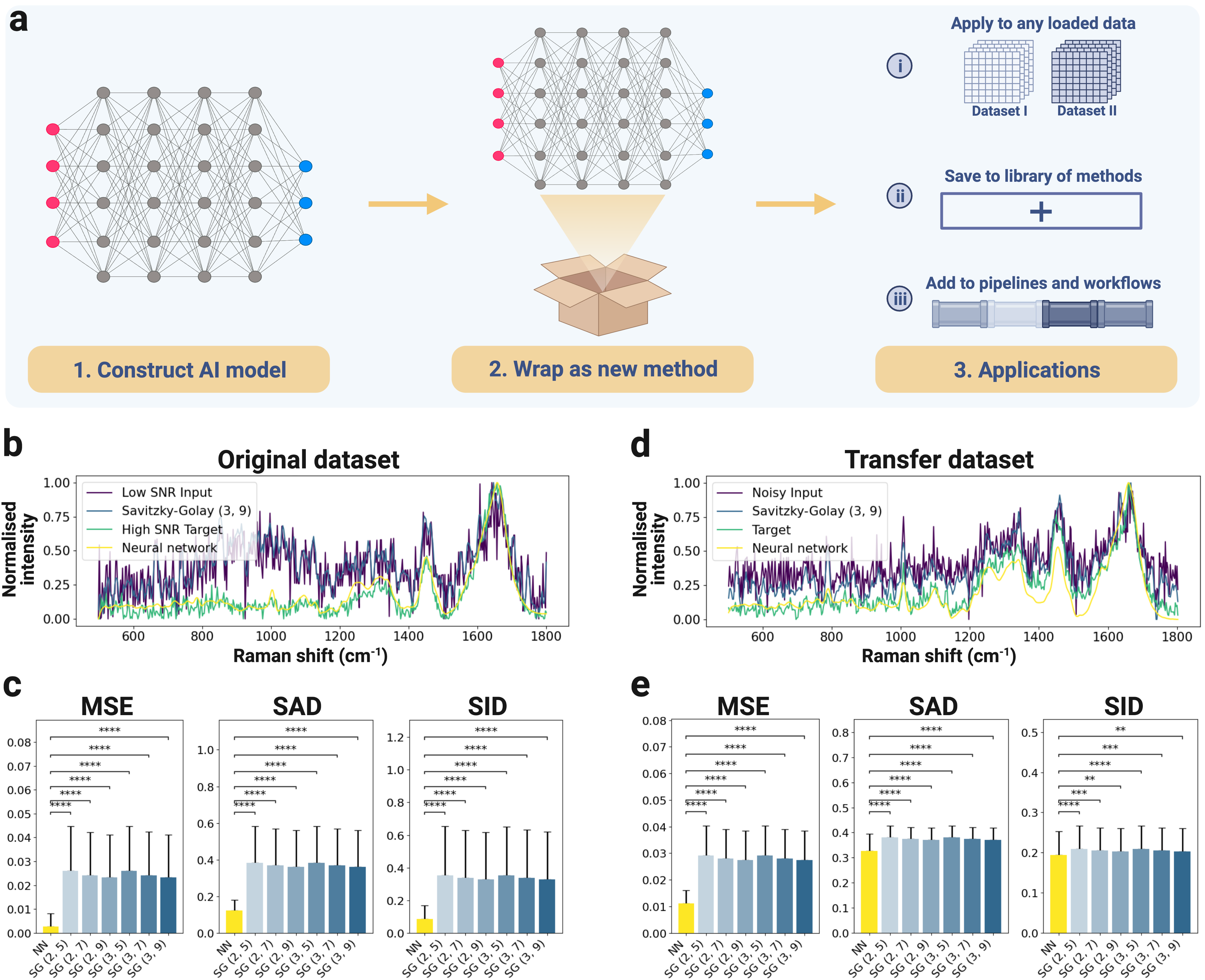}
    \caption{\textbf{\package{} interfaces with AI/ML Python frameworks to create new methods for RS analysis.} \textbf{a,} \package{} allows users to incorporate AI/ML models seamlessly into pipelines created within the platform. \textbf{b-c,} A pre-trained 1D ResUNet deep-learning denoiser~\cite{deeper} is integrated as a preprocessing module within \package{} to investigate its performance against the Savitzky-Golay (SG) filter~\cite{savgol}. \textbf{b, } Denoising of a spectrum from~\cite{deeper}, where the low-SNR (purple) is the input and the high-SNR (green) is the target. The data is denoised with a SG filter of polynomial order 3 and kernel size 9, SG(3, 9) (blue), and with the implemented deep-learning denoiser (yellow).  \textbf{c, } The results on the test set from \cite{deeper} ($n=12694$) show that the deep-learning denoiser outperforms six SG filters across three performance metrics (MSE, SAD, SID). Error bars represent one standard deviation around the sample mean. Statistical significance measured with a two-sided Wilcoxon signed-rank test with adjustment for multiple comparisons based on Benjamini-Hochberg correction \cite{Benjamini_Hochberg} (* $P < 0.05$, ** $P < 0.01$, *** $ P < 0.001$, **** $P < 0.0001$).
    \textbf{d-e,} Same analysis on unseen data from \cite{kallepitis2017quantitative} ($n=1600$). The input (purple) corresponds to data contaminated with added noise and the target (green) to the original data. In this case, the deep-learning denoiser only shows an improvement for MSE.
    }
    \label{fig:integration1}
\end{figure}

First, \package{} allows the seamless integration of standard Python AI/ML methods (e.g., from \emph{scikit-learn} \cite{sklearn}, \emph{PyTorch} \cite{pytorch} and \emph{tensorflow} \cite{tensorflow}) as tools for RS analysis (Fig. \ref{fig:integration1}a).
As an illustration, we use \package{} to construct a deep learning denoising procedure based on the one-dimensional ResUNet model - a fully convolutional UNet neural network with residual connections~\cite{deeper}. To do this, we simply wrap within \package{} the pre-trained neural network (trained on spectra from MDA-MB-231 breast cancer cells, available at \url{https://github.com/conor-horgan/DeepeR}) as a custom denoising method. Once wrapped, the denoiser is automatically compatible with the rest of \package{} and can be readily employed for different applications. For instance, we replicate the results in \cite{deeper}, and show in Fig.~\ref{fig:integration1}b-c that the application of this deep-learning denoiser to the low signal-to-noise ratio (SNR) test set from \cite{deeper} consistently outperforms the commonly-used Savitzky-Golay filter \cite{savgol}, as quantified by various metrics also coded within \package{} (e.g., mean squared error (MSE), spectral angle distance (SAD) \cite{sad} and spectral information divergence (SID) \cite{sid}). 
Applying this pipeline to new data only involves changing the data source. 
Taking advantage of this transferability, we test the denoiser on unseen volumetric Raman data from another cell line (THP-1 \cite{kallepitis2017quantitative}), with added Gaussian noise (see SI). 
In this case, Fig. \ref{fig:integration1}d-e shows improved performance especially according to the MSE metric, which is dependent on normalisation, but with lower significance according to scale-invariant and information-theoretic metrics, also available in \package{}. This example emphasises the importance of incorporating robust validation criteria within data analysis workflows.

\begin{figure}
    \centering
    \includegraphics[width=1\textwidth]{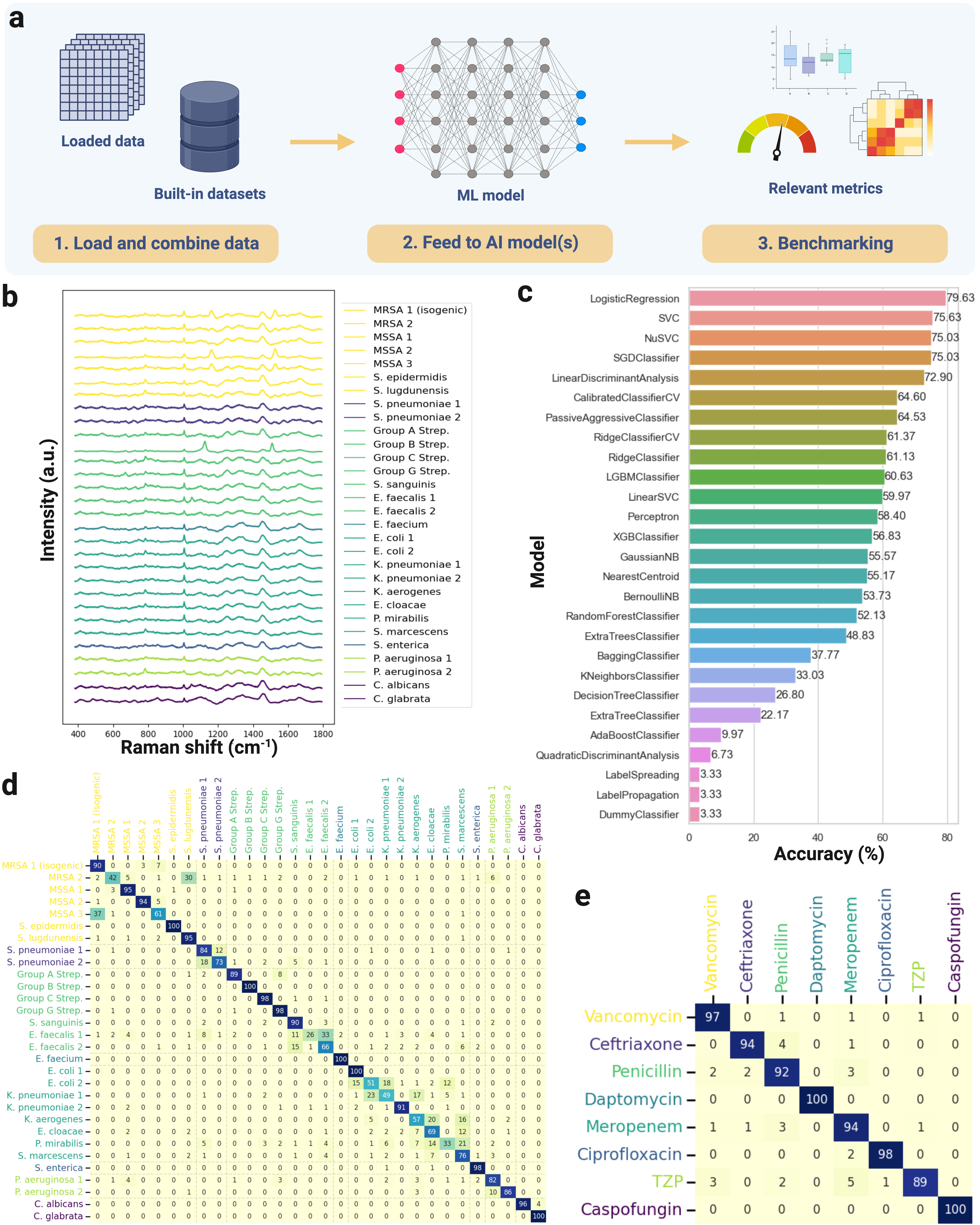}
    \caption{\textbf{\package{} as a suite for algorithmic development and benchmarking.} \textbf{a,}  Data representations in \package{} are compatible with the Python AI/ML ecosystem, allowing data flow from \package{} to \emph{scikit-learn} \cite{sklearn}, \emph{PyTorch} \cite{pytorch}, \emph{tensorflow} \cite{tensorflow}, etc. \package{} is also equipped with standard datasets and relevant metrics to support model development and validation. 
    \textbf{b-e} Benchmarking ML classification models on the task of bacteria identification from Raman spectra~\cite{bacteria}. \textbf{b,} Mean Raman spectra for all bacterial species provided (100 spectra per species). Spectra are min-max normalised to the range 0--1 for visualisation purposes. \textbf{c,} Benchmarking results of 28 ML models.
    The best accuracy was achieved by the logistic regression classifier. \textbf{d-e,} Confusion matrices for the best species-level (\textbf{d}) and antibiotic-level (\textbf{e}) classifier with accuracies of 79.63\% and 94.63\%, respectively. 
    }
    \label{fig:integration2}
\end{figure}

Secondly, the data management backbone of \package{} ensures a direct data flow to the rest of the Python ecosystem, i.e., data can be loaded, preprocessed, and analysed in \package{} and then exported to conduct further modelling and analysis elsewhere (Fig. \ref{fig:integration2}a). As an example application, we perform AI-based bacteria identification using Raman measurements \cite{bacteria} from 30 bacterial and yeast isolates (Fig. \ref{fig:integration2}b). After loading and exploring the spectra with \package{}, we interface the data with the \emph{lazypredict} Python package \cite{lazypredict} and benchmark 28 different ML classification models (including logistic regression, support vector machines and decision trees) on the task of predicting the species from the spectrum. The models were trained on a high-SNR dataset (100 spectra per isolate) and tested on an unseen high-SNR testing set of the same size. Our benchmarking analysis in Fig. \ref{fig:integration2}c finds logistic regression as the best-performing model, achieving a classification accuracy of 79.63\% on the species-level classification task (Fig. \ref{fig:integration2}d), and 94.63\% for antibiotic treatment classification (Fig. \ref{fig:integration2}e).

To further assist validation against previous results, \package{} provides access to a library of curated datasets, which can be integrated into analysis and benchmarking workflows. This lays the foundation for a common repository of RS data and reduces barriers to data access, especially for ML teams with limited access to RS instruments \cite{luo2022deep}. The dataset library in \package{} already includes data loaders for Raman data from bacterial species \cite{bacteria}, cell lines \cite{deeper, kallepitis2017quantitative}, COVID-19 samples \cite{yin2021efficient, Yin2020}, multi-instrument Surface Enhanced Raman Spectroscopy (SERS) measurements of adenine samples \cite{fornasaro2020surface}, wheat lines \cite{csen2023differentiation}, minerals \cite{rruff}, and will continue to be expanded. 

\section*{Discussion}\label{sec12}
In this paper, we have introduced \package{} - a computational framework for integrative Raman spectroscopic data analysis. \package{} offers a comprehensive collection of tools for spectroscopic analysis designed to systematise the RS data analysis life cycle, reducing typical overheads of analysis workflows and improving methodological standardisation.
The package also lays the foundations of a common repository of standardised methods, protocols and datasets, which users can readily access and exploit within the \package{} framework to conduct different benchmarking studies.
Furthermore, \package{} is fully compatible
with frameworks for data science and machine learning in Python, thereby facilitating the adoption and validation of advanced AI technologies for next-generation RS analysis. Lastly, we remark that, while our focus here has been on Raman spectroscopy, many of the tools in \package{} are of broad applicability to other vibrational spectroscopy techniques, including infrared (IR) spectroscopy.

\section*{Methods}\label{sec11}
\subsection*{Installation}
\package{} has been deposited in the Python Package Index (\url{https://pypi.org/project/ramanspy}). This means it can be directly installed via the common package installer \emph{pip} for Python:
\begin{lstlisting}[language=bash]
  pip install ramanspy
\end{lstlisting}

To access the functionalities of the package after installation, users only need to import \package{} in their Python scripts. One can import the whole package:
\begin{lstlisting}[language=Python, breaklines]
import ramanspy
# or import ramanspy as rp
\end{lstlisting}
or individual modules or methods:
\begin{lstlisting}[language=Python, breaklines]
# individual modules
from ramanspy import load, preprocessing

# individual methods
from ramanspy.analysis.unmix import NFINDR
\end{lstlisting}

\subsection*{Core infrastructure}

\bmhead{Data management}
Data in \package{} is represented by a set of custom data container classes based on scalable, computationally efficient array programming \cite{numpy}, which correspond to different spectroscopic modalities. This includes the generic \textit{SpectralContainer} class, as well as the more specialised \textit{Spectrum}, \textit{SpectralImage} and \textit{SpectralVolume} classes representing on single-point spectra (1D), imaging data (3D), volumetric data (4D) respectively. These classes define data-specific information and behaviour in the background to allow a smooth, user-friendly experience, regardless of the data of interest.

The containers can be initialised by providing the corresponding intensity data, the spectral axis (in cm$^{-1}$) and other relevant (meta) data, which will become properties of the constructed object. For instance: 
\begin{lstlisting}[language=Python, breaklines]
raman_spectrum = ramanspy.Spectrum(intensity_data, spectral_axis, *args, **kwargs)
raman_image = ramanspy.SpectralImage(intensity_data, spectral_axis, *args, **kwargs)
\end{lstlisting}

Once created, data containers can be manipulated, visualised, saved and loaded as needed using the built-in tools in \package{}. 

Note that for the most part, users would not need to manually populate these containers. Instead, they can take advantage of the data loading functionalities that \package{} provides.

\bmhead{Data loading}
To support data loading, \package{} offers easy-to-use data loaders compatible with experimental Raman spectroscopic data from a range of instrumental vendors in the area. These loaders - available within \emph{ramanspy.load} - automatically parse relevant data files and return the appropriate spectral container. As an example, users can load MATLAB files exported from WITec's ProjectFOUR/FIVE software using the following command:
\begin{lstlisting}[language=Python, breaklines]
raman_object = ramanspy.load.witec(<PATH>)
\end{lstlisting}

A full list of the data loaders built into \package{} is available as part of the documentation of the package at \url{https://ramanspy.readthedocs.io/en/latest/loading.html}.

Raman data can also be loaded via established data-loading tools in Python. For instance, one can use \emph{pandas'} \textit{csv} loader to load a spectrum from a \textit{.csv} file with two columns storing the intensity data and the spectral axis by using:
\begin{lstlisting}[language=Python, breaklines]
import pandas as pd

data = pd.read_csv(csv_filename)
raman_spectrum = ramanspy.Spectrum(data["<intensity_column>"], data["<axis_column>"])
\end{lstlisting}

\bmhead{Spectral preprocessing}
Preprocessing logic in \package{} is defined by the \textit{PreprocessingStep} class, which defines most of the necessary preprocessing infrastructure in the background to ensure a smooth, data-agnostic experience via a single point of contact specified through their \textit{apply()} method.

Yet, as with data loading, for the most part, users are not expected to use this class to manually implement and optimise such preprocessing methods themselves. Instead, the \package{} package provides a comprehensive toolbox of ready-to-use preprocessing methods, which users can access, customise and employ to compile a wide variety of preprocessing procedures. These preprocessing procedures are given as predefined classes within \emph{ramanspy.preprocessing} which extend the \textit{PreprocessingStep} class. To use these built-in methods, users need to create an instance of the selected technique. For instance:
\begin{lstlisting}[language=Python, breaklines]
denoiser = ramanspy.preprocessing.denoise.SavGol(*args, **kwargs)
baseline_corrector = ramanspy.preprocessing.baseline.ASLS(*args, **kwargs)
normaliser = ramanspy.preprocessing.normalise.MaxIntensity(*args, **kwargs)
\end{lstlisting}
Note that \package{} offers full control over relevant parameters, which can be supplied during initialisation via the \textit{*args} and \textit{**kwargs} arguments.

As the methods inherit all operational logic defined within the parent \textit{PreprocessingStep} class, they can be directly accessed and used on any data loaded in the framework through their \textit{apply()} method:
\begin{lstlisting}[language=Python, breaklines]
preprocessesd_objects = denoiser.apply(<spectral object or collection of spectral objects>)
preprocessesd_objects = baseline_corrector.apply(<spectral object or collection of spectral objects>)
\end{lstlisting}

A full list of the methods for spectral preprocessing built into \package{} is available as part of the documentation of the package at \url{https://ramanspy.readthedocs.io/en/latest/preprocessing.html}.

If needed, users can also incorporate any in-house method into \package{} by manually creating instances of the \textit{PreprocessingStep} class which wrap the given method. This can be done as follows:
\begin{lstlisting}[language=Python, breaklines]
def preprocessing_func(intensity_data, spectral_axis, *args, **kwargs):
    # Preprocess intensity_data and spectral_axis
    ...

    return updated_intensity_data, updated_spectral_axis

# wrapping the function together with the relevant *args and **kwargs
custom_preprocessing_method = ramanspy.preprocessing.PreprocessingStep(preprocessing_func, *args, **kwargs)
\end{lstlisting}

Then, the custom preprocessing method is fully compatible with the rest of \package{}'s functionalities and out-of-the-box applicable to any data integrated within the package via its \textit{apply()} method:
\begin{lstlisting}[language=Python, breaklines]
custom_preprocessing_method.apply(<spectral object or collection of spectral objects>)
\end{lstlisting}

Note that this class structure implies that these instances can then be  saved (e.g. as \textit{pickle} files) and, therefore, reused and shared as required afterwards.

\bmhead{Spectral analysis}
As with preprocessing classes, users can access any built-in analysis method (available within the \emph{ramanspy.analysis} sub-module) by creating an object instance of the corresponding class (again - with full control over relevant parameters) as follows:
\begin{lstlisting}[language=Python, breaklines]
nmf = ramanspy.analysis.decompose.NMF(*args, **kwargs)
kmeans = ramanspy.analysis.cluster.KMeans(*args, **kwargs)
unmixer = ramanspy.analysis.unmix.NFINDR(*args, **kwargs)
\end{lstlisting}

Once created, instances can be similarly accessed via their \textit{apply()} method on any data loaded in \package{}.
\begin{lstlisting}[language=Python, breaklines]
cluster_maps, cluster_centres = kmeans.apply(<spectral object or collection of spectral objects>)
abundance_fractions, endmemebrs = unmixer.apply(<spectral object or collection of spectral objects>)
\end{lstlisting}

A full list of the methods for spectral analysis built into \package{} is available as part of the documentation of the package at \url{https://ramanspy.readthedocs.io/en/latest/analysis.html}.

\bmhead{Visualisation}
The \package{} package also provides various visualisation tools available within the \emph{ramanspy.plot} sub-module. As an example, one can plot spectra using the \textit{spectra} function:
\begin{lstlisting}[language=Python, breaklines]
ramanspy.plot.spectra(<spectra or collection of spectra>)
ramanspy.plot.show()  # or plt.show() after import matplotlib.pyplot as plt
\end{lstlisting}
Note that these functions are highly customisable. This can be done by providing relevant parameters to control the plot generation, as well as through \textit{matplotlib}'s customisation workflow.
\begin{lstlisting}[language=Python, breaklines]
import matplotlib.pyplot as plt

plt.figure(figsize = (5, 5))
ax = ramanspy.plot.spectra(<spectra or collection of spectra>, title="<str>", label="<str or list[str]>")
ax.set_ylabel("<str>")  # adding a label to the y-axis
plt.show()  # or ramanspy.plot.show()
\end{lstlisting}

A full list of the methods for data visualisation built into \package{} is available as part of the documentation of the package at \url{https://ramanspy.readthedocs.io/en/latest/plot.html}.

\subsection*{Preprocessing pipelines}
Pipelining behaviour is defined by the \textit{Pipeline} class in \package{}, which ensures that pipelines are accessible, simple-to-use and fully compatible with the rest of \package{}. 

\bmhead{Creating a custom preprocessing pipeline}
To assemble a preprocessing pipeline, one simply needs to stack relevant methods (built-in or custom) into the intended order of execution. For instance:
\begin{lstlisting}[language=Python, breaklines]
preprocessing_pipeline = ramanspy.preprocessing.Pipeline([
    ramanspy.preprocessing.denoise.SavGol(*args, **kwargs),
    ramanspy.preprocessing.baseline.ASLS(*args, **kwargs),
    ramanspy.preprocessing.normalise.MaxIntensity(*args, **kwargs),
    custom_preprocessing_method(*args, **kwargs)  # custom in-house method
])
\end{lstlisting}

Constructed pipelines can then be applied exactly as single methods via their \textit{apply()} method to any data loaded within \package{}.
\begin{lstlisting}[language=Python, breaklines]
preprocessesd_objects = preprocessing_pipeline.apply(<spectral object or collection of spectral objects>)
\end{lstlisting}

As pipelines in \package{} are objects, they can also be directly saved in a convenient file format, such as \textit{pickle} files. As such, they can then be reloaded, reused and shared as needed.

\bmhead{Access a predefined preprocessing pipeline}
\package{} also provides a collection of built-in preprocessing pipelines. 
To access them, one can select the desired protocol from \emph{ramanspy.preprocessing.protocols} as follows:
\begin{lstlisting}[language=Python, breaklines]
preprocessing_pipeline = ramanspy.preprocessing.protocols.PROTOCOL_X
\end{lstlisting}
A pre-defined \textit{Pipeline} instance will be returned, which can similarly be employed directly through its \textit{apply()} method.

A full list of the protocols for spectral preprocessing built into \package{} is available as part of the documentation of the package at \url{https://ramanspy.readthedocs.io/en/latest/preprocessing.html#established-protocols}.

\subsection*{AI integration}

\bmhead{Integrate AI methods into \package{}}
To integrate new techniques for spectral preprocessing and analysis, users can take advantage of the extensible architecture of \package{} and wrap models and algorithms into custom classes. For instance, one can create a new denoiser method based on a \emph{PyTorch} model for denoising by simply creating a function, which defines how the model can be used to preprocess a generic intensity data array, and then wrapping the method within a \textit{PreprocessingStep} instance.
\begin{lstlisting}[language=Python, breaklines]
def nn_preprocesing(intensity_data, wavenumber_axis):
    intensity_data = v.reshape(-1, intensity_data.shape[-1])
    output = model(torch.Tensor(intensity_data).unsqueeze(1)).cpu().detach().numpy()
    output = np.squeeze(output).reshape(intensity_data.shape)

    return output, wavenumber_axis
    
nn_denoiser = ramanspy.preprocessing.PreprocessingStep(nn_preprocesing)
\end{lstlisting}

Integrated methods are automatically rendered fully compatible with the rest of \package{}'s functionalities in the background, so one can simply use the \textit{apply()} method of the constructed denoiser to preprocess any data loaded within \package{} as any built-in preprocessing class.

\bmhead{Export data from \package{} to AI frameworks}
The data management core of \package{} allows a direct interface with the entire Python ecosystem, including frameworks for statistical modelling, machine learning and deep learning. To do that, users can simply feed relevant data from \package{} to functions and tools they want to use elsewhere. For instance, one can pass the intensity data stored in a spectral container to a specific model from the \textit{scikit-learn} \cite{sklearn} framework for statistical and ML modelling directly via their \textit{fit()} method:
\begin{lstlisting}[language=Python, breaklines]
model.fit(spectral_container.spectral_data)
\end{lstlisting}

\subsection*{Datasets}
To access the Raman spectroscopic datasets available in \package{}, users can employ custom data-loading methods built into \package{} under \emph{ramanspy.datasets}. These would automatically parse the relevant data into the corresponding spectral container. For instance, one can load the bacteria data from \cite{bacteria} using the following function:
\begin{lstlisting}[language=Python, breaklines]
data_container, labels = ramanspy.datasets.bacteria(dataset="train", <PATH>)
\end{lstlisting}

Note that, depending on where each dataset was deposited and the license it was deposited under, some of these methods will automatically download the given dataset, whereas others may require the manual download of the data. Users are pointed to the documentation of each method for instructions on how to properly load each dataset.

A full list of the datasets built into \package{} is available as part of the documentation of the package at \url{https://ramanspy.readthedocs.io/en/latest/datasets.html}.

\subsection*{Metrics}
Users can likewise readilyaccess relevant spectroscopic metrics, such as MSE, SAD and SID, from \emph{ramanspy.metrics}. These can be used to measure the similarity between spectra by using the respective method:
\begin{lstlisting}[language=Python, breaklines]
ramanspy.metrics.SID(spectrum_I, spectrum_II)
\end{lstlisting}

A full list of the metrics built into \package{} is available as part of the documentation of the package at \url{https://ramanspy.readthedocs.io/en/latest/metrics.html}.


\newpage

\bibliography{sn-bibliography}

\section*{Declarations}

\subsection*{Data availability}
All data used in this article are previously published open-access data that have been deposited by the respective authors online. Instructions on how to access, download and load the datasets provided in \package{} are available in the documentation at \url{https://ramanspy.readthedocs.io/en/latest/datasets.html}.

\subsection*{Code availability}
The codebase of \package{} is open-source and hosted on GitHub at \url{https://github.com/barahona-research-group/RamanSPy}. The package can be installed via \emph{pip} using '\emph{pip install ramanspy}'. Documentation, including detailed tutorials and examples, is available at \url{https://ramanspy.readthedocs.io}. The scripts used to produce the analysis results presented in this paper are also provided as executable Jupyter Notebook examples at \url{https://github.com/barahona-research-group/RamanSPy/tree/3dd2c1e09420c5ac473a72ebd6ed06a91c30a85c/paper_reproducibility} and as part of the documentation of \package{} at \url{https://ramanspy.readthedocs.io/en/latest/auto_examples/index.html}.

\subsection*{Acknowledgments}
D.G. is supported by UK Research and Innovation [UKRI Centre for Doctoral Training in AI for Healthcare grant number EP/S023283/1]. 
S.V.P. gratefully acknowledges support from the Independent Research Fund Denmark (0170-00011B). 
R.X. and M.M.S. acknowledge support from the Engineering and Physical Sciences Research Council (EP/P00114/1 and EP/T020792/1).
A.F.G. acknowledges support from the Schmidt Science Fellows, in partnership with the Rhodes Trust.
M.M.S. acknowledges support from the Royal Academy of Engineering Chair in Emerging Technologies award (CiET2021\textbackslash\textbackslash94).
M.B. acknowledges support by the EPSRC under grant EP/N014529/1, funding the EPSRC Centre for Mathematics of Precision Healthcare at Imperial College London, and under grant EP/T027258/1.
The authors thank Dr Akemi Nogiwa Valdez for proofreading and data management support. 

Figures were created with BioRender (\url{www.biorender.com}).


\newpage
~\newpage

\section*{Supplementary information}

Several of our examples are based on data from \cite{kallepitis2017quantitative} which provided volumetric RS scans across 4 distinct THP-1 cell lines. Here, we only used the first scan (scan \textit{'001'}). 

\subsection*{Cell phenotyping via spectral unmixing.} 
The raw THP-1 data from \cite{kallepitis2017quantitative} 
used for the spectral unmixing procedure in Fig.~\ref{fig:core_example} was re-exported as MATLAB files from the \textit{WITec Project FIVE} software. The MATLAB files were then loaded into \package{} followed by spectral preprocessing with a protocol consisting of: (1) spectral cropping to the $700-1800$cm$^{-1}$ region; (2) cosmic rays removal with the algorithm in \cite{whitaker2018simple}; (3) denoising with a Savitzky-Golay filter polynomial order 3 and kernel size 7 \cite{savgol}; (4) baseline correction with asymmetric least squares \cite{eilers2005baseline}; and (5) Global MinMax normalisation to the interval $[0,1]$. 

After preprocessing, we performed spectral unmixing in \package{} using N-FINDR \cite{nfindr} (number of endmembers set to 5) and FCLS \cite{fcls}. We concluded the analysis by visualising the results corresponding to the top 4 endmembers. 


\subsection*{Preparing THP-1 data for deep learning denoising.}
The denoising analysis on the data in Fig.~\ref{fig:integration1}d-e was performed on the middle depth layer (fifth layer out of 10) of the THP-1 volumetric scan from \cite{kallepitis2017quantitative}. This layer consisted of a $40 \times 40$ image scan, i.e., 1600 spectra. To be consistent with the original paper~\cite{deeper}, we conducted exactly the same preprocessing protocol described there. Namely, we utilised the \textit{WITec Project FIVE} software to crop the data to the region $500-1800$cm$^{-1}$, followed by baseline correction using the ‘\textit{shape}’ method with $\alpha=500$.

To assess the performance of the deep learning denoiser, we created `low-SNR spectra' by adding Gaussian noise to the original spectra.
Each spectrum was MinMax-normalised to the range 0--1 and Gaussian noise with a standard deviation $\sigma=0.15$ was added. This resulted in spectra of similar noise levels to those in \cite{deeper}. 
These noisy samples were used as the input to the model and the uncontaminated data was taken as ground-truth targets. 

We then MinMax-normalised each spectrum (both inputs and targets) and compared the performance of the neural network denoiser against six Savitzky-Golay filters \cite{savgol}. To make all models comparable, and to correct for potential artefacts of how the model was trained originally in \cite{deeper}, all denoising metrics were computed after MinMax-normalising the denoised outputs of each denoiser to the range 0--1 again.


\end{document}